\documentclass[doublecol]{epl2} 

\usepackage{graphicx}
\usepackage{dcolumn}
\usepackage{bm}
\usepackage{color}
\usepackage{appendix}
\usepackage{siunitx}
\usepackage{ulem}
\usepackage{xfrac}
\usepackage{amsmath}
\usepackage{amssymb}
\usepackage{ulem}

\newcommand{\vk}{von K\'{a}rm\'{a}n~}

\newcommand*\Laplace{\mathop{}\!\mathbin\bigtriangleup}
\newcommand{\starr}[1]{#1^{*}}

\title{Experimental observation of spontaneous temperature fluctuations in turbulent flows}

\author{G. Prabhudesai\inst{1,2} \and S. Perrard\inst{1,3} \and F. P\'{e}tr\'{e}lis\inst{1} \and S. Fauve\inst{1}}
\shortauthor{G. Prabhudesai \etal}

\institute{                    
  \inst{1} Laboratoire de Physique de l'\'{E}cole Normale Sup\'{e}rieure, CNRS, PSL Research University, Sorbonne Universit\'{e}, Universit\'{e} de Paris, F-75005 Paris, France \\
  \inst{2} University of California Los Angeles, Los Angeles, California 90095, USA \\
  \inst{3} PMMH, CNRS, ESPCI Paris, Université PSL, Sorbonne Université, Université Paris-Cité, 75005 Paris, France
}

\abstract{
Even in the absence of externally applied temperature gradients, spontaneously generated temperature fluctuations arise in turbulent flows. We experimentally study these fluctuations in a closed \vk swirling flow of air at Mach number of order $10^{-3}$, whose boundaries are maintained at a constant temperature. We observe intermittent peaks of low temperature correlated with pressure drops within the flow and show that they are caused by vorticity filaments. The measured ratio of temperature to pressure fluctuation agrees with the prediction based on adiabatic cooling within vortex cores. This experimental study shows that although the Mach number of the flow is small, there exist regions within the flow where compressible effects cannot be discarded in the equation for temperature and locally dominate the effect of viscous dissipation. }

\begin{document}

\maketitle

\section{Introduction} 

Temperature fluctuations spontaneously generated in turbulent flows have not been studied experimentally so far, in contrast to advection of temperature in the presence of an externally applied temperature gradient \cite{corrsin1951spectrum,obukhov1970structure}. Two processes are responsible for spontaneously generated temperature fluctuations. First, viscous dissipation of the flow kinetic energy generates heat, as recently studied in detail using numerical simulations \cite{de2013temperature,bos2014temperature,bos2015}. Fluctuations of temperature result from bursts of dissipation that occur at small scales and in a strongly intermittent way in space and time. These fluctuations could therefore provide additional information about small scale intermittency of turbulent flows. There exists a second mechanism other than dissipative processes, which subsists in the limit of a perfect fluid with conserved entropy. It is related to heating or cooling induced by pressure fluctuations generated by the flow. The equation for temperature in the absence of any externally applied large scale gradients is given by \cite{monin2007statistical},

\begin{align}
	\rho c_{p}\frac{DT}{Dt} = \bigg( \frac{\rho\nu}{2} \bigg)\Sigma^{2} + \sigma\frac{Dp}{Dt} + \alpha\Laplace{T} \label{eqn:1a}
\end{align}

where $\sfrac{D}{Dt} = (\sfrac{\partial}{\partial t} + \bm{u}\cdot\nabla)$ and $\Sigma = \big[(\nabla \bm{u}) + (\nabla \bm{u})^{T}\big]$ respectively denote the material derivative and strain rate tensor, and, $\Sigma^{2} = \Sigma_{ij}\Sigma_{ji}$. The quantities of density, specific heat capacity at constant pressure, kinematic viscosity, thermal conductivity and coefficient of thermal expansion are denoted by $\rho$, $c_{p}$, $\nu$, $\alpha$ and $\beta$ respectively. We denote the non-dimensionalized coefficient of thermal expansion by $\sigma = \beta T$. The first two terms on the right hand side of eqn.~\ref{eqn:1a} correspond to two sources of spontaneously generated temperature fluctuations; viscous dissipation and compressible heating/cooling driven by pressure fluctuations. As demonstrated by Bayly \etal \cite{bayly1992density}, both the source terms in eqn.~\ref{eqn:1a} are retained in the limit of incompressible flow with the quantities replaced by $\starr{\rho}$, $\starr{c_{p}}$, $\starr{\nu}$, $\starr{\alpha}$, $\starr{\beta}$ and $\starr{\sigma}=\starr{\beta}\starr{T}$. The superscripted asterisk denotes the values of the corresponding quantities measured in the absence of flow. We define the Reynolds number as $Re = UL/\starr{\nu}$ and Peclet number as $Pe = \starr{\rho} \starr{c_{p}} UL/\starr{\alpha}$ where $U$, $L$ are the characteristic velocity and length scales of the flow. In the limit of incompressible flow, eqn.~\ref{eqn:1a} then implies that when $\starr{\sigma} Re \gg 1$, temperature fluctuations are mostly driven by pressure fluctuations\footnote{This condition is obtained on comparing the magnitudes of the first two terms on the right hand side of eqn.~\ref{eqn:1a} which gives $(\starr{\sigma}Dp/Dt) \Big/ (\starr{\rho}\starr{\nu}\Sigma^{2}) \sim (\starr{\sigma}\starr{\rho}U^{3}/L) \Big/ (\starr{\rho}\starr{\nu}U^{2}/L^{2}) = \starr{\sigma}Re$, using $p \sim \starr{\rho} U^{2}$ and $\Sigma \sim U/L$. }. Additionally, in this case, the fluid will undergo adiabatic process if the condition of $Pe = Pr Re \gg 1$ is also met, where the Prandtl number is defined as $Pr = \starr{\rho} \starr{c_{p}} \starr{\nu}/\starr{\alpha}$. When $\starr{\sigma} Re \ll 1$, they are mostly generated by viscous dissipation. The above analysis, though, is not as straightforward for turbulent flows where a wide range of scales coexist and interact, with the flow exhibiting a range of Reynolds numbers across scales. We can nonetheless ascertain that the temperature fluctuations at small scales $(Re \ll 1)$ would be governed by viscous dissipation and at large scales $(Re \gg 1)$ by pressure fluctuations. We note that this second mechanism has been discarded in numerical simulations of temperature equation, even though it should certainly be taken into account at large Reynolds numbers, for instance in astrophysical or geophysical flows. At the laboratory scale, adiabatic cooling is observed within strong coherent vortices such as in the Ranque tube \cite{ranque} or in \vk swirling flows between co-rotating disks \cite{labbe1996study,labbe1996ecoulements} where a significant temperature drop can be observed in the vortex core \cite{labbe2014filamentary}. Generating a \vk turbulent swirling flow in air ($\starr{\sigma} \approx 1$, Pr = 0.71) between counter-rotating disks, we show that even in the absence of an externally driven vortex and at moderate values of the Reynolds number, these compressible events contribute significantly to the temperature fluctuations that are generated spontaneously by the flow.

\section{Experimental set-up and results}  

\begin{figure}[htbp!]
	\centering
	\includegraphics[height=9cm,width=0.9\linewidth]{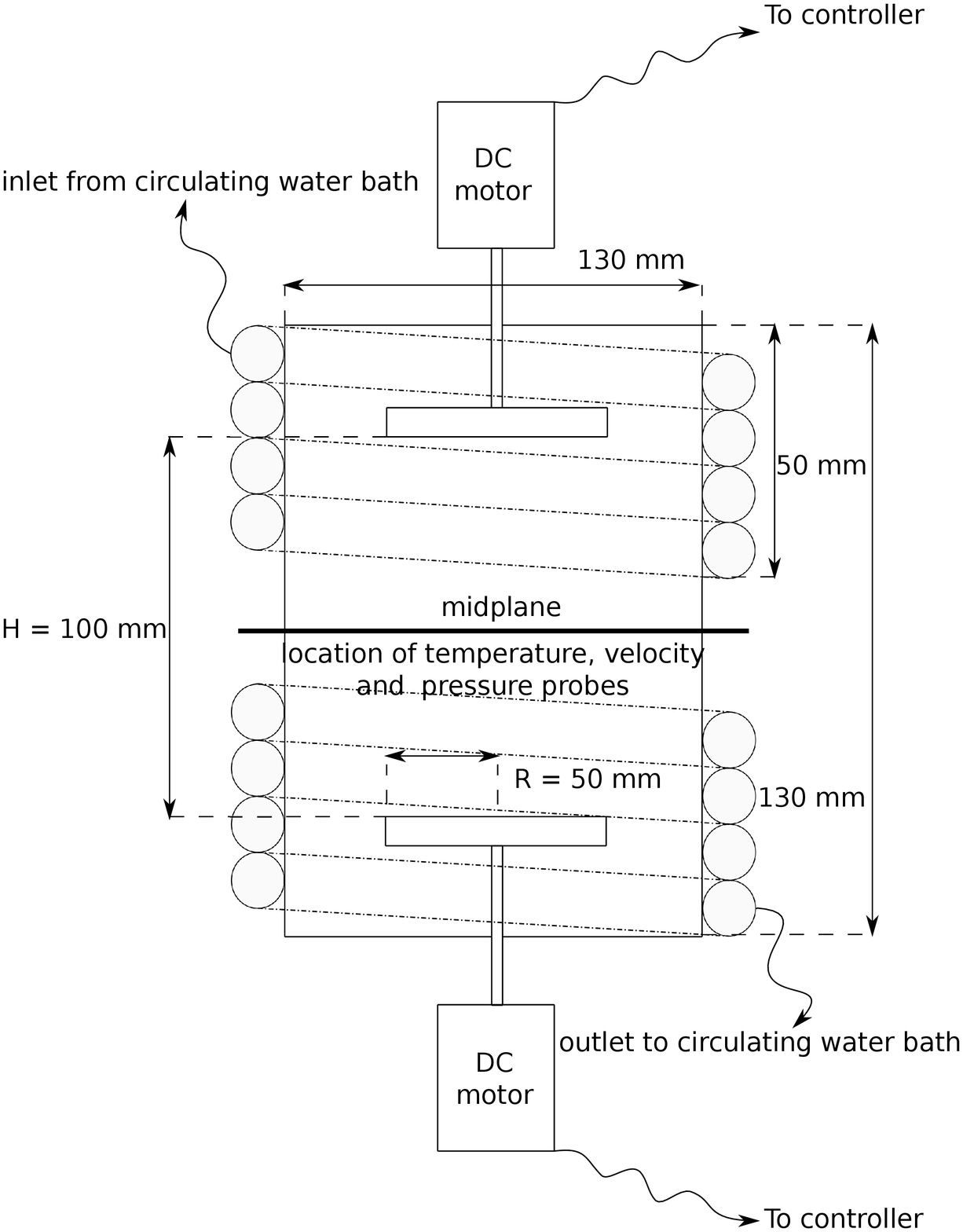}
	\includegraphics[height=6.5cm,width=0.8\linewidth]{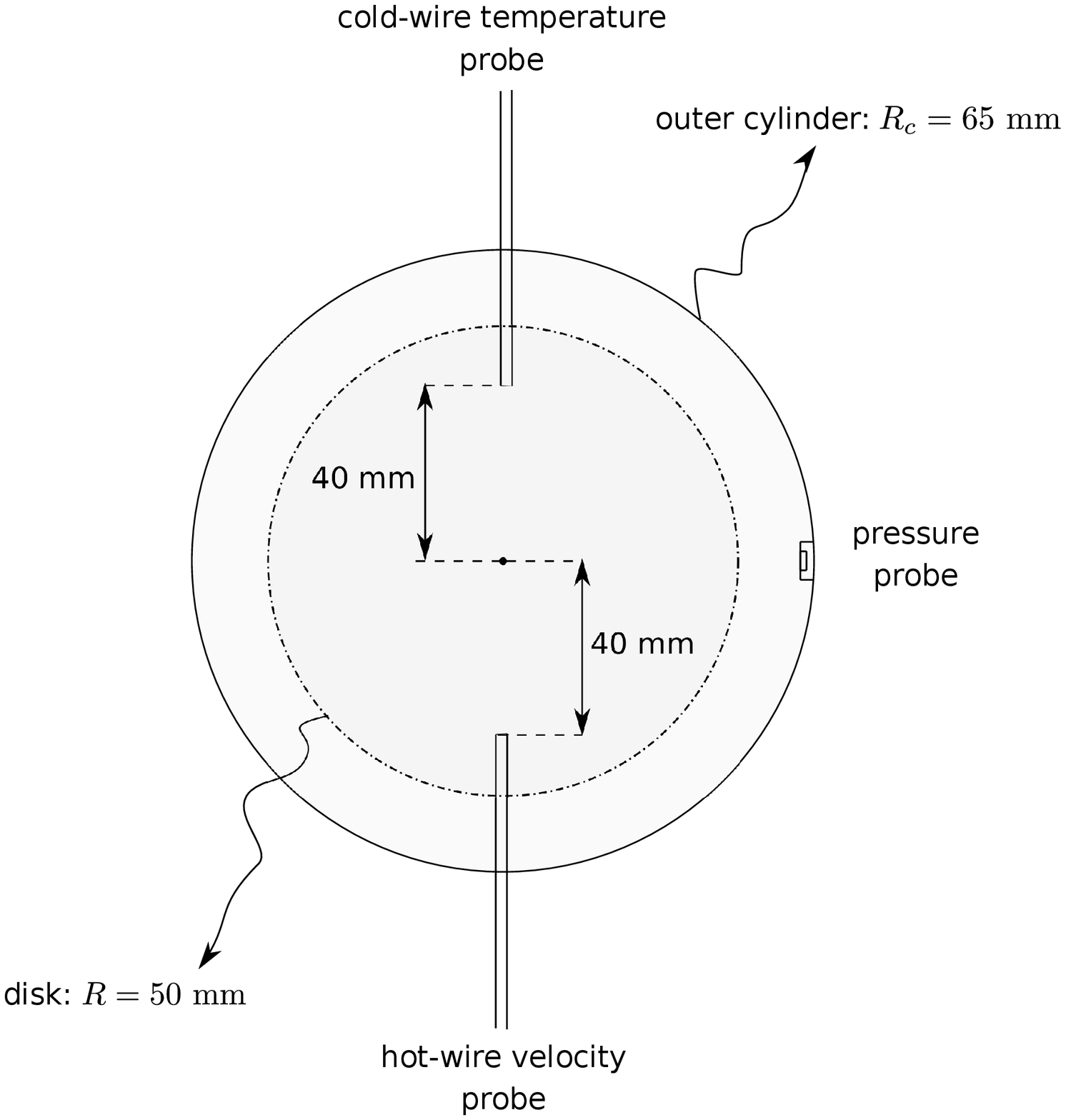}
	\caption{Sketch of the \vk flow configuration driven by two counter-rotating disks. Temperature of the boundaries is maintained constant using a circulating water bath. The pressure, velocity and temperature probes are all placed in the midplane.}
	\label{fig:1}
\end{figure}

We generate a \vk swirling flow in air between two counter-rotating disks in a closed cylinder of diameter $130$ mm made out of copper of $2 \ \si{\mm}$ thickness (see fig. \ref{fig:1} top). Copper has been chosen for its high thermal conductivity to minimize temperature gradients along the boundaries of the experiment. The top and the bottom faces of the cylinder are closed with circular copper plates of the same thickness as the cylinder. In order to maintain fixed the temperature at the boundaries, a copper tubing of outer diameter $10 \ \si{\mm}$ is welded to the cylindrical part of the container. The inlet and the outlet of the copper tubing are connected to a circulating water bath with thermostat control which maintains the temperature of the water at a given fixed value within $\pm 0.01 \ \si{\K}$. 

We use two loop-controlled brushless DC motors with the angular velocity $\Omega$ ranging from of $0 - 2000$ rotations per minute (rpm). Each of the motor drives a disk with four curved blades which generate a strong flow inside the cylindrical cavity. The thickness of the disks and the height of the blades are $7.5 \ \si{\milli\meter}$. Holes drilled on the surface of the cylinder provide access to the probes: one cold-wire temperature probe (Dantec 55P31)\footnote{Diameter of the filament $d_{c} = 1 \ \mu$m, response time $\tau_{r}^{c} = 0.5$ ms, resolution of $0.2$ mK, uncertainty of $\pm 1$ mK.}, one 1D hot-wire velocity probe (Dantec 55P16) and one acceleration compensated piezoelectric pressure probe (PCB 103B02)\footnote{Diaphragm diameter $d_{p} = 2.5$ mm, frequency range at $\pm 3$ dB is $1$ Hz - $7$ kHz, resolution of $0.15$ Pa, uncertainty of $\pm 1$ Pa.}. All probes are placed in the midplane, with the pressure probe flushed to the wall as shown in the top-view sketch of fig.~\ref{fig:1} bottom. The mean power consumption per unit mass by the turbulent flow $\langle \epsilon \rangle$ is measured from the power required by the motors to maintain the flow, ranging from $0$ to $500$ m$^2$/s$^{3}$. This gives the Kolmogorov length scale $\eta = \big(\nu^{3}/\langle \epsilon \rangle\big)^{1/4} \geq 50$~\si{\micro\metre} and the corresponding Kolmogorov timescale $\tau_{\eta} = \langle \epsilon \rangle^{-1/3}\eta^{2/3} \geq 0.2$ ms with $\nu = 1.5 \times 10^{-5}$~m$^2$/s the kinematic viscosity of air at room temperature. We evaluate the Taylor microscale $\lambda = \big( 15 \nu u_{rms}^{2}/\langle \epsilon \rangle \big)^{1/2} \geq 0.8$~\si{\milli\metre}, the corresponding Taylor timescale $\tau_{\lambda} = \langle \epsilon \rangle^{-1/3}\lambda^{2/3} \geq 1$ ms and the Taylor microscale based Reynolds number at the highest rotation rate $Re_{\lambda} \sim O(10^{2})$ using standard estimates for homogeneous and isotropic turbulent flows~\cite{tennekes1972first}. The cold-wire temperature probe should thus be able to resolve turbulent scales as small as $\lambda$. Data is acquired from the probes at a sampling frequency of $5$ kHz using a NI DAQ system. Two point spatial correlation of velocity is evaluated in a different set of experiments using two 1D hot-wire probes moved radially about the center axis in the midplane. An exponential decay is observed giving an integral length scale of $l_{0} = 5 \ \text{mm}$. The Mach number defined as $Ma = u_{rms}/\starr{c}$ reaches a maximum value of $4 \times 10^{-3}$ for the current setup where $u_{rms}$ is the RMS of velocity fluctuations and $\starr{c}$ is the isentropic speed of sound in the fluid at rest.

\begin{figure}[htbp!]
	\centering
	\includegraphics[height=6cm,width=\linewidth]{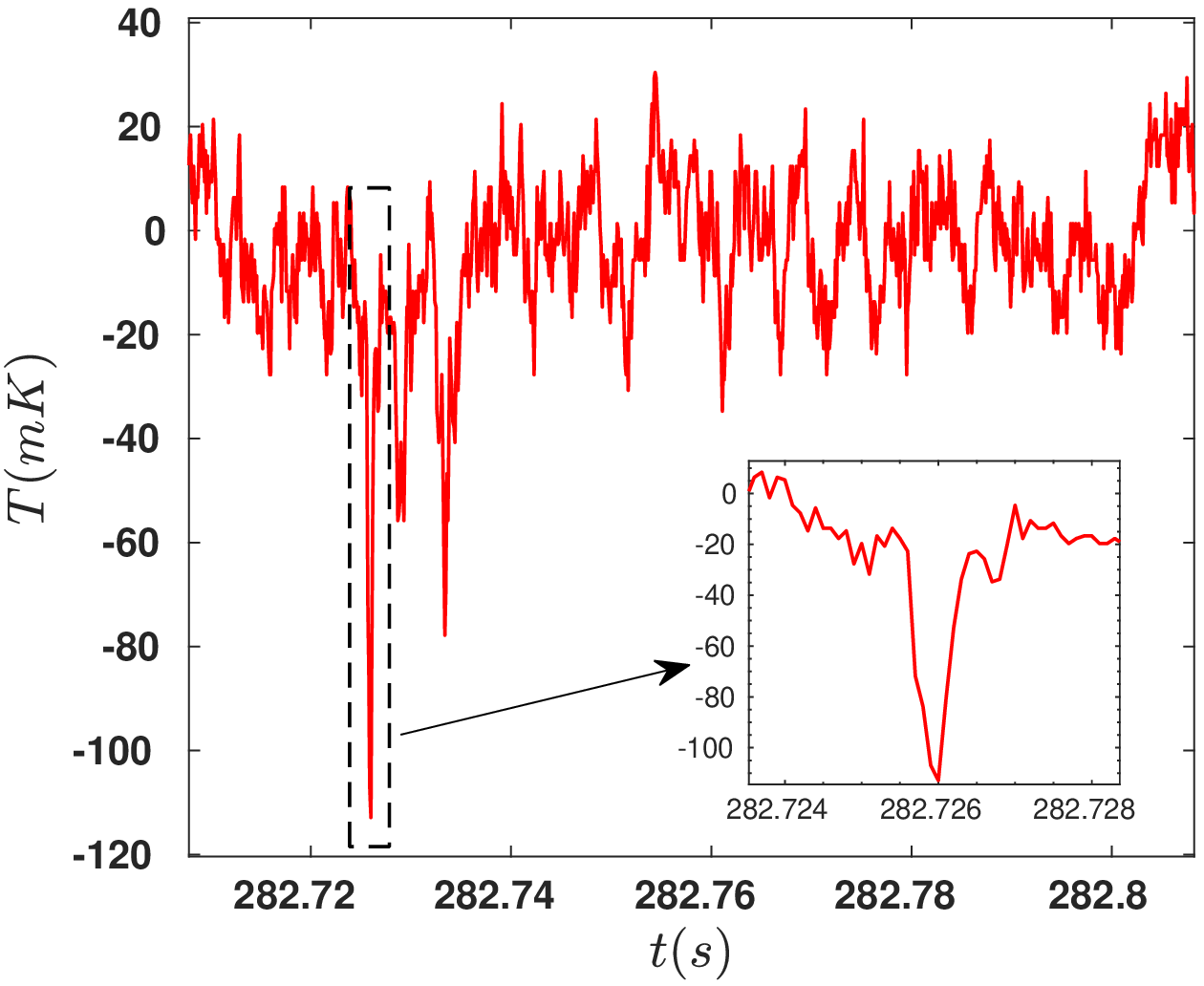}
	\includegraphics[height=6cm,width=\linewidth]{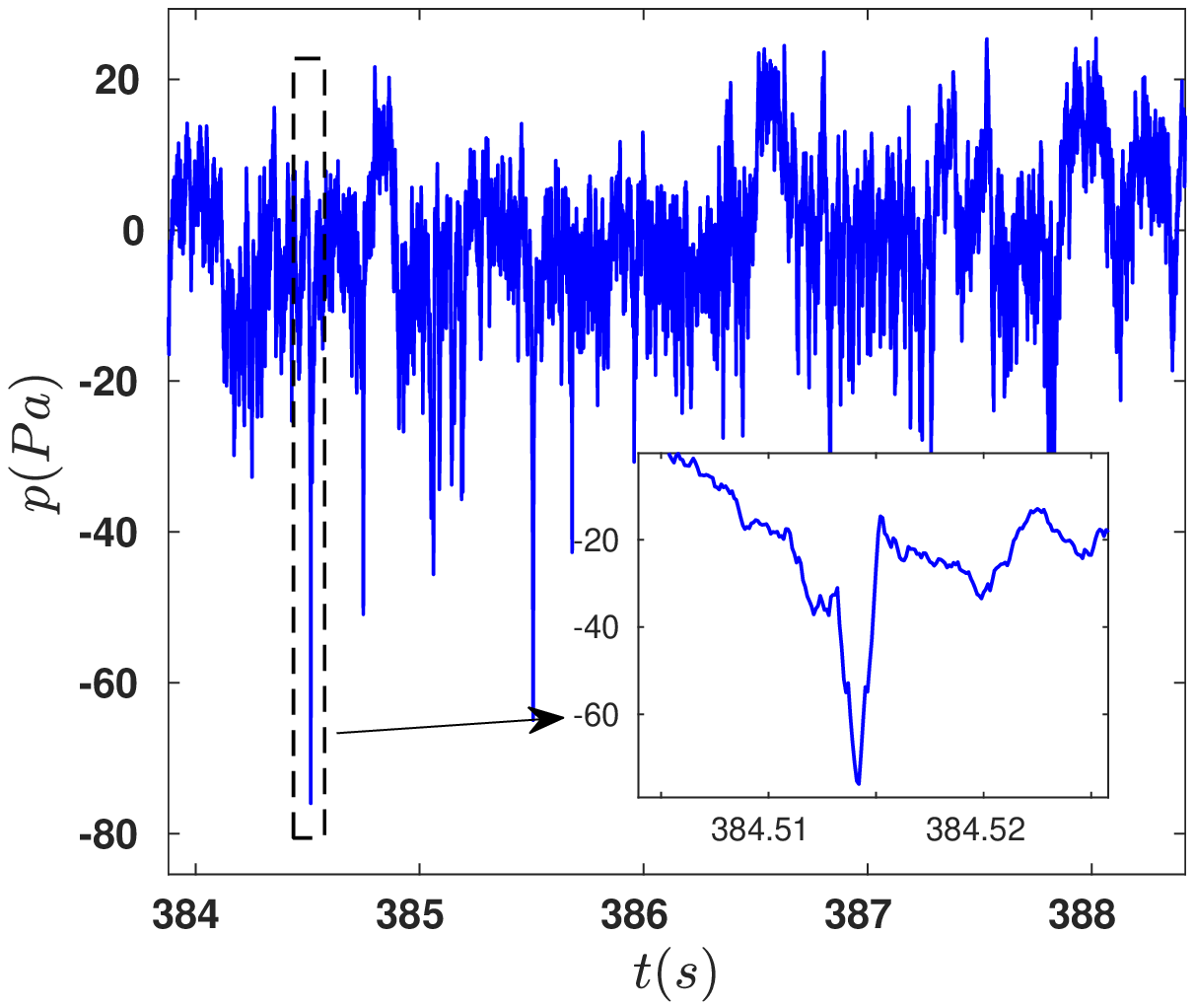}
	\caption{(a) Time series of temperature fluctuations measured in the bulk. (b) Time series of pressure fluctuations measured at the boundary. Inset figures show the zoom of one drop observed in temperature and pressure fluctuations. The drops are intermittent, i.e., they are observed to occur randomly in time.}
	\label{fig:2}
\end{figure}

\begin{figure}[htbp!]
	\centering
	\includegraphics[height=6cm,width=\linewidth]{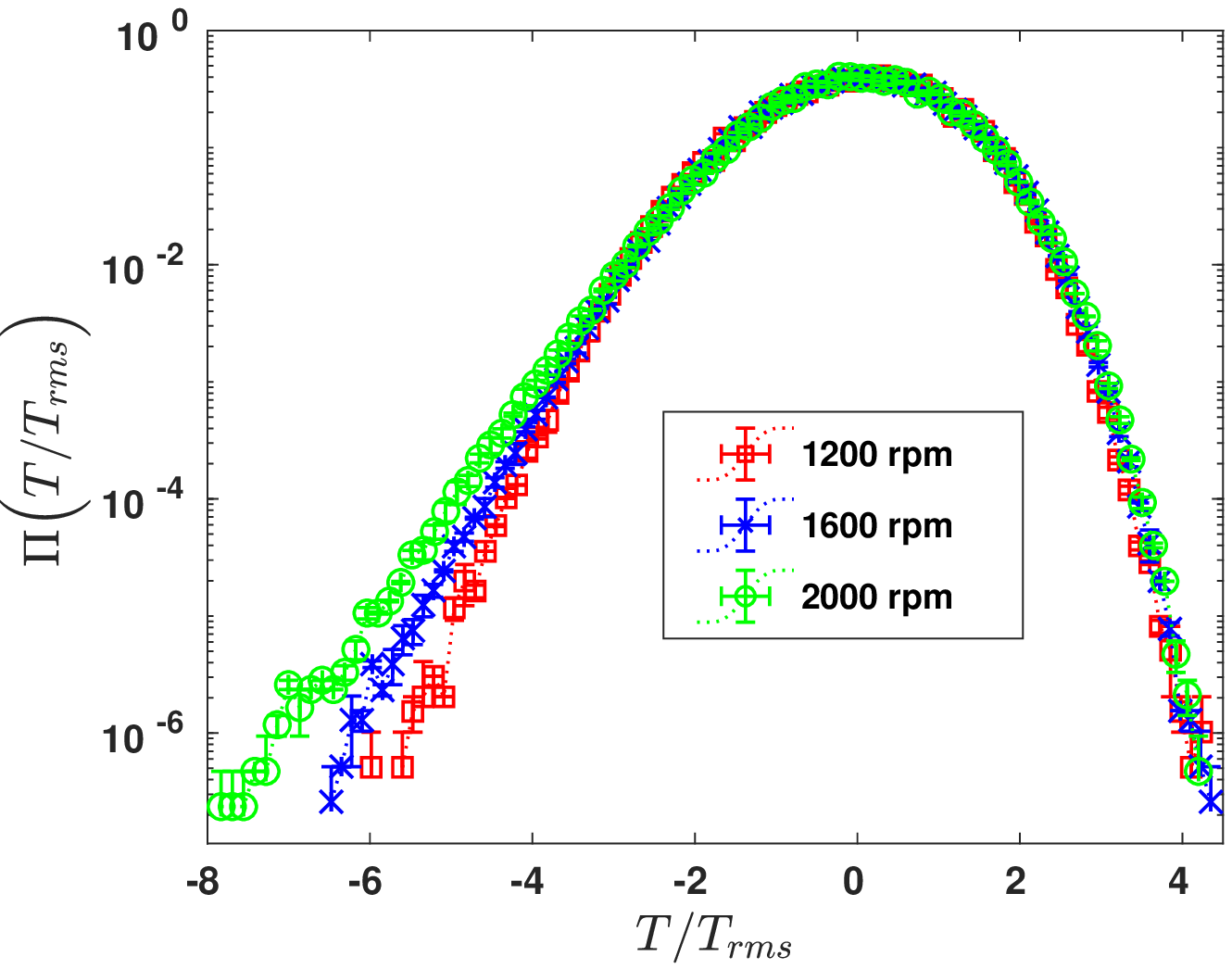}
	\includegraphics[height=6cm,width=\linewidth]{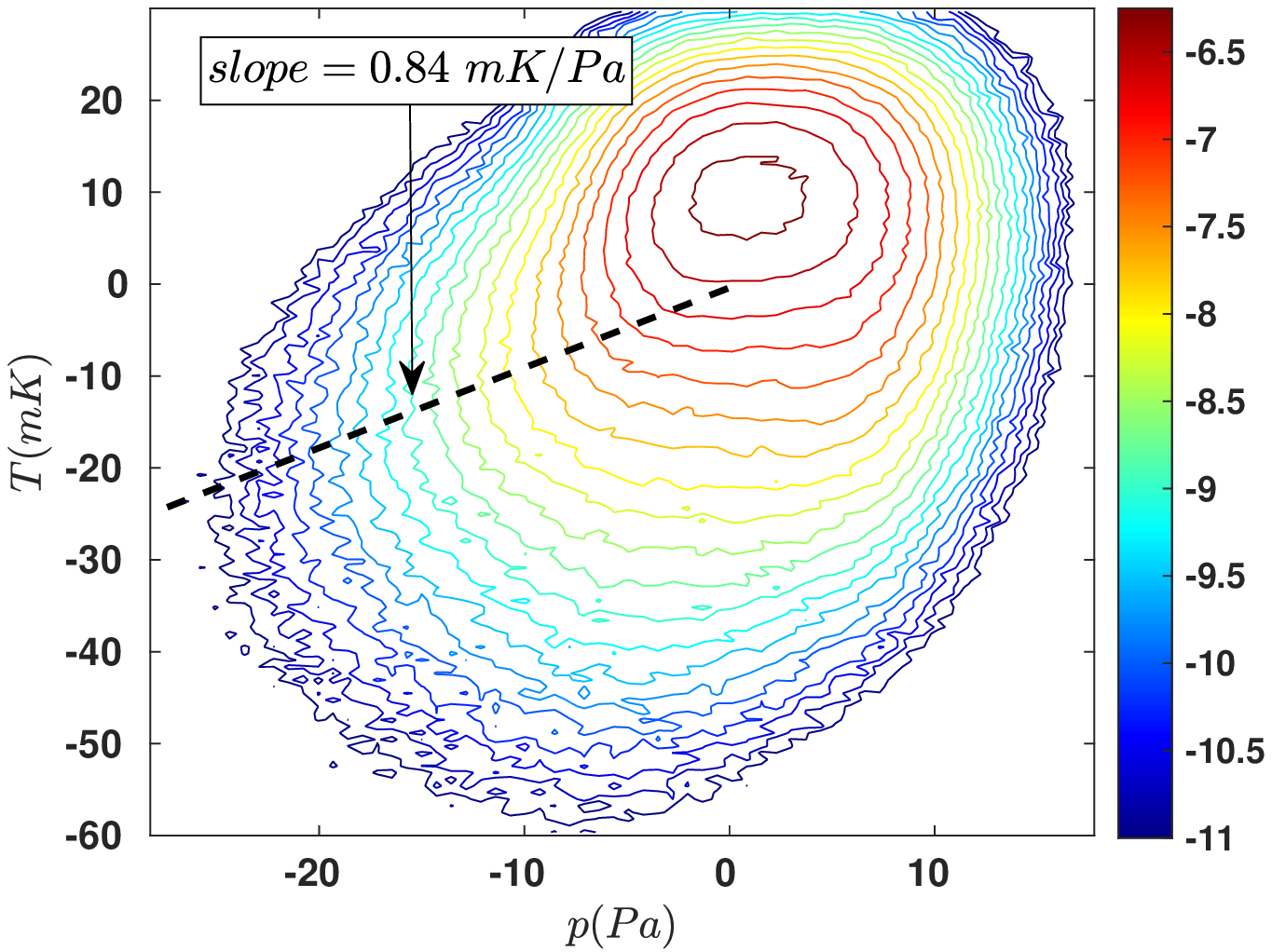}
	\caption{(a) PDFs of temperature fluctuations normalized by their respective RMS values for $\Omega = 1200$ rpm (red), $1600$ rpm (blue) and $2000$ rpm (green). (b) The joint PDF of pressure and temperature fluctuations when the two probes are $6$ mm apart for $\Omega = 2000~\text{rpm}$. The isocontours of probability are plotted on a natural logarithmic scale. The black dashed line has a slope of $0.84 \ mK/Pa$.}
	\label{fig:3}
\end{figure}

Fig.~\ref{fig:2}a and \ref{fig:2}b show the time series of temperature and pressure fluctuations about their mean values, obtained for $\Omega = 2000$ rpm. Sharp drops are observed in both signals, examples of which are shown in the respective inset figures. The intermittent pressure drops in a turbulent flow have been fairly well-studied and so is the physical phenomenon associated with it. These have been attributed to vorticity filaments \cite{douady1991direct,fauve1993pressure,cadot1995characterization,dernoncourt1998experimental,ishihara2003spectra} and were shown to result in the behaviour at low frequency of the pressure power spectrum \cite{abry1994analysis}. The observed amplitude of pressure drops is of $O(\starr{\rho}R^{2}\Omega^{2})$ which is in agreement with previous experimental observations \cite{cadot1995characterization}. Similar to the pressure signal, we observe intermittent temperature drops much larger than the standard deviation of the temperature fluctuations. They correspond to the low exponential tail of the temperature PDFs (fig.~\ref{fig:3}a). Previous studies \cite{meneveau1991multifractal,yeung2012dissipation,buaria2019extreme} have shown that the process of viscous dissipation is intermittent, with the PDFs of $\epsilon$ being positively skewed. As observed from eqn.~\ref{eqn:1a}, this would result in peaks in temperature and the process of viscous dissipation thus cannot explain the observed drops.

One immediate question that arises is whether the drops observed in pressure and temperature signals are correlated and result from the same type of structures in the turbulent flow. To check that, we evaluate the joint PDFs between pressure and temperature signals. To do so, we place the temperature probe close to the pressure probe ($d \approx 6 \ \si{\mm}$ apart) near the boundary. The joint PDF between pressure and temperature is defined as,

\begin{eqnarray}
	\Pi(p_{0},T_{0},\Delta t) dp_{0}\,dT_{0} =\phantom{aaaaaaaaaaaaaaaaa} \\ \text{prob}\Big(p(t)\in [p_{0},p_{0}+dp_{0}],T(t+\Delta t)\in [T_{0},T_{0}+d T_{0}] \Big)&\nonumber
	\label{eqn:temp1}
\end{eqnarray}
for a time lag $\Delta t$ between the pressure and temperature signals. The correlation between pressure and temperature shows a maximum for a non zero time lag $\Delta t_{max} \sim 2$ \si{\milli\second} which can be attributed to the advection of fluctuations between the two probes by a mean flow. This gives a characteristic advection velocity of $u_{c} = d/\Delta t_{max} = 3$ \si{\meter/\second} which is consistent with the velocity associated with disk rotation of $O(R\Omega)$. Fig.~\ref{fig:3}b shows the joint PDF \ $\Pi(p_{0},T_{0},\Delta t = \Delta t_{max})$ obtained for a rotation rate of $\Omega = 2000$ rpm. We observe that the joint PDF is skewed towards  negative values of both pressure and temperature fluctuations. We conclude that the vorticity filaments are responsible not only for the distribution of pressure drops but also for the temperature drops\footnote{If the pressure and temperature signals were independent, the isocontours of joint PDFs would be skewed along negative $p$-axis and negative $T$-axis. This would be due to the PDFs of $p$ and $T$ individually being negatively skewed but without correlation between them.}. From eqn.~\ref{eqn:1a}, in the limit of $\starr{\sigma}Re \gg 1$ and $Pe \gg 1$, and for the fluid being an ideal gas, the amplitudes of pressure ($\Delta p$) and temperature ($\Delta T$) satisfy the adiabatic relation,

\begin{eqnarray}
	\frac{\Delta T}{\Delta p} =  \bigg(\frac{\gamma-1}{\gamma}\bigg)\bigg(\frac{\starr{T}}{\starr{p}}\bigg) \approx 0.84 ~\text{mK/Pa}
	\label{eqn:temp2},
\end{eqnarray}

\begin{figure}
\centering
\includegraphics[height=6cm,width=\linewidth]{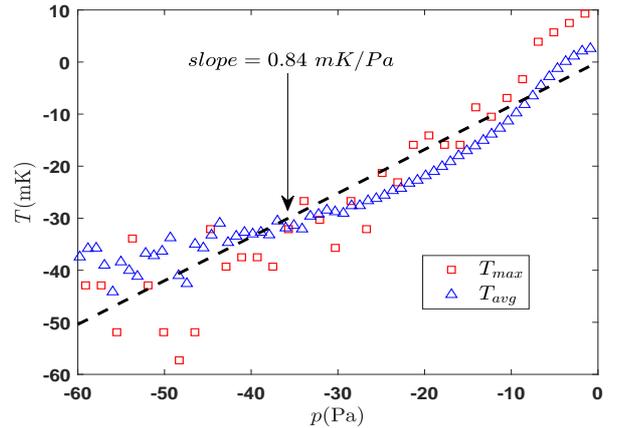}
\caption{$T_{max}$ (the temperature at which the joint PDF between $p$ and $T$ is maximized, $\textcolor{red}{\small{\Box}}$) and $T_{avg}$ (the conditional average of $T$ on $p$, $\textcolor{blue}{\small{\triangle}}$) vs $p$ for $\Omega = 2000$ rpm. The black dashed line shows the prediction for adiabatic process (eqn.~\ref{eqn:temp2}).}
\label{fig:4}
\end{figure}

where $\gamma \approx 1.4$ for air at normal temperature and pressure. For a given value of $p$, we can define the temperatures $T_{max}(p)$ and $T_{avg}(p)$. The value of temperature that maximizes the conditional PDF $\Pi(T|p,\Delta t_{max})$ is denoted by $T_{max}(p)$, whereas, $T_{avg}(p) = \int_{-\infty}^{+\infty} T \ \Pi(T|p,\Delta t_{max}) dT$ denotes the conditional average of $T$ on $p$. Fig.~\ref{fig:4}a shows the values of these two temperatures extracted from fig. \ref{fig:3}b. We observe that for negative values of pressure fluctuations, $T_{max}$ and $T_{avg}$ depend linearly on $p$ with a slope of $0.84 ~\text{mK/Pa}$ as predicted by eqn.~\ref{eqn:temp2}. From this observation we conclude that concurrent negative fluctuations in pressure and temperature are associated with adiabatic cooling. It is worth noting that the temporal width of the drops fluctuates following an exponential distribution for large values, with a characteristic time of $5$ ms for pressure drops and $1$ ms for temperature drops. The difference results from the finite size of pressure probe, as studied by Abry \etal\cite{abry1994analysis} using different probe sizes. In our experiments, the dimensions of the pressure and temperature probes are $2.5$ \si{\milli\metre} and $1$ \si{\micro\metre} respectively. Since the physical dimension of the measuring element of the temperature probe is smaller than any length scale associated with the turbulent flow, it turns out to be a better probe to access the internal structure of vorticity filaments. 

From previous studies \cite{cadot1995characterization,jimenez1998characteristics}, we can estimate the mean value of $\sigma^* Re_f$ and $Pe_f$ in the core of vorticity filaments. The typical velocity scale of the vorticity filaments is $u_{rms}$ and their width lies between $\eta$ and $\lambda$. Thus for the current experimental setup, we obtain $5 < \starr{\sigma}Re_{f} < 100$ and $3.6 < Pe_{f} < 71$ corresponding to $\starr{\sigma}Re_{f} > 1$ and $Pe_{f} > 1$, for which the adiabatic cooling/heating dominates over viscous dissipation, as we observe experimentally.

\section{Conclusion}~
We have provided experimental evidence that the vorticity filaments in turbulent flows contribute to temperature fluctuations even for small Mach numbers of order $10^{-3}$ through adiabatic effects. This is evidenced by the correlation between the drops observed in both pressure and temperature fluctuations. It should be noted that our observations are not limited to gas. If one considers the case of water at $20 ^{\circ}C$ for which $\starr{\sigma} \approx 0.06$ and $Pr \approx 7$, temperature fluctuations would be driven by adiabatic heating/cooling when $Re \gg 16$, a value easily attained in water flows. The magnitude of temperature fluctuations thus generated would be $70$ times smaller than those in air for same velocity as deduced from eq. \ref{eqn:1a} but should be measurable at larger velocities since the pressure drops scale as the square of velocity. 

Vorticity filaments being characteristic of all turbulent flows, this process should be considered in a variety of contexts. Not only  when studying temperature fluctuations, but also  when studying a field that is coupled to temperature, possible examples of which are quite numerous. In most cases, the field under consideration, say $n$ to give an example, will display relative fluctuations equal to the ones of temperature, so that $dn/n=dT/T$. We measure here temperature drops of relative amplitude of $O(10^{-2})$ percent. As they are quadratic in velocity, large relative fluctuations can easily be observed with a moderate increase in velocity. One example is the scattering of acoustic or electromagnetic waves due to temperature fluctuations in turbulent flows \cite{tatarski2016wave}, in which case $n$ denotes the refractive index of the medium.

Another example relevant at the astrophysical scale is that of interstellar medium which is known to be turbulent \cite{larson1979stellar,Larson_1981} and displays intermittency \cite{falgarone1990signature}. Later studies (see for example \cite{falgarone1995intermittency,godard2014chemical,momferratos2014turbulent}) showed the effects of intermittent turbulent structures, in particular dissipative structures on the chemistry of interstellar medium. Our work shows that pressure-driven adiabatic temperature fluctuations can dominate over the effect of viscous dissipation and thus calls for studying the effect of these fluctuations on the chemistry of interstellar medium. 

In the domain of cloud physics, an open question is how rain forms in warm cumulus on a short time scale (less than one hour) \cite{rogers}.  Recent investigations have highlighted the effect of turbulent fluctuations on the rapid growth of cloud droplets across the size gap from $10$ to $50$ $\mu$m. Several mechanisms have been considered: enhanced geometric collision rates by air turbulence \cite{falkovich2002}, effect of turbulence on collision efficiencies, effect of stochastic fluctuations \cite{wang2006,prabhakaran2020role}. The present study shows that condensation induced by adiabatic cooling which is essentially the working principle of cloud chambers, should also be taken into account. How adiabatic cooling observed in vorticity filaments affects the formation of cloud droplets and whether this contributes to the efficiency of precipitation in warm rain process  deserves further studies.

\acknowledgments
This work has been supported by the Agence nationale de la recherche (Grant No. ANR-17-CE30-0004), CEFIPRA (Project 6104-1) and by CNES (action 6291).

\bibliographystyle{eplbib} 
\bibliography{biblio}

\end{document}